\documentclass[conference,10pt]{IEEEtran}

\usepackage{psfrag}
\usepackage{amsmath}
\usepackage{amssymb}
\usepackage{amsfonts}
\usepackage{graphicx}
\usepackage{float}
\usepackage{graphics}
\usepackage{graphicx}
\usepackage{theorem}
\usepackage{euscript}

\newcommand{\abs}[1]{\left\vert#1\right\vert}

\theoremstyle{plain}
\newtheorem{thm}{Theorem}

\newtheorem{prop}[thm]{Proposition}

\begin{document}

\title{Cellular Systems with Full-Duplex Amplify-and -Forward Relaying and Cooperative Base-Stations}

\author{\authorblockN{Oren Somekh\authorrefmark{1}, Osvaldo Simeone\authorrefmark{2},
H. Vincent Poor\authorrefmark{1}, and Shlomo Shamai
(Shitz)\authorrefmark{3}}
\authorblockA{\authorrefmark{1}
Department of Electrical Engineering, Princeton University,
Princeton, NJ 08544, USA}
\authorblockA{\authorrefmark{2}
CWCSPR, Department of Electrical and Computer Engineering, NJIT,
Newark, NJ 07102, USA}
\authorblockA{\authorrefmark{3}
Department of Electrical Engineering, Technion, Haifa 32000,
Israel}
\thanks{This research was supported by the OIF Marie-Curie 6th program and the NSF CNS-06-25637 grants}}


\maketitle

\begin{abstract}
In this paper the benefits provided by multi-cell processing of
signals transmitted by mobile terminals which are received via
dedicated relay terminals (RTs) are assessed. Unlike previous
works, each RT is assumed here to be capable of full-duplex
operation and receives the transmission of adjacent relay
terminals. Focusing on intra-cell TDMA and non-fading channels, a
simplified uplink cellular model introduced by Wyner is
considered. This framework facilitates analytical derivation of
the per-cell sum-rate of multi-cell and conventional single-cell
receivers. In particular, the analysis is based on the observation
that the signal received at the base stations can be interpreted
as the outcome of a two-dimensional linear time invariant system.
Numerical results are provided as well in order to provide further
insight into the performance benefits of multi-cell processing
with relaying.
\end{abstract}

\section{Introduction}
Techniques for provision of better service and coverage in
cellular mobile communications are currently being investigated by
industry and academia. In this paper, we study the combination of
two cooperation-based technologies that are promising candidates
for such a goal, extending previous work in
\cite{Simeone-Somekh-BarNess-Spagnolini-WCOM06}
\cite{Simeone-Somekh-Barness-Spagnolini-IT06}. The first is
relaying, whereby the signal transmitted by a mobile terminal (MT)
is forwarded by a dedicated relay terminal (RT) to the intended
base station (BS) \cite{Lin-Hsu-Infocom00} (see also
\cite{Pabst-Walke-Schultz-Herhold-Yanikomeroglu-Mukherjee-Viswanathan-Lott-Sirwas-Falconer-Fettweis-COMMAG04}
for a more recent account). The throughput of such hybrid networks
has recently been studied in the limit of asymptotically many
nodes
\cite{Zemlianov-Veciana-JSAC05}\cite{Liu-Liu-Towsely-INFOCOM03}.
Moreover, information theoretic characterization of related
single-cell scenarios has been reported in
\cite{Kramer-Gastpar-Gupta-IT05}. The second technology of
interest here is multi-cell processing (MCP), which allows the BSs
to jointly decode the received signals, equivalently creating a
distributed receiving antenna array
\cite{Zhou-Zhao-Xu-Yao-COMMAG03}. The performance gain provided by
this technology within a simplified cellular model was first
studied in \cite{Wyner-94}\cite{Hanly-Whiting-Telc-1993}, and then
extended to include fading channels by \cite{Somekh-Shamai-2000},
under the assumption that BSs are connected by an ideal backbone
(see
\cite{Shamai-Somekh-Zaidel-JWCC-2004}\cite{Somekh-Simeone-Barness-Haimovich-Shamai-BookChapt-07}
for surveys on MCP).

Recently, the interplay between these two technologies has been
investigated for amplify-and-forward (AF) and decode-and-forward
(DF) protocols in \cite{Simeone-Somekh-BarNess-Spagnolini-WCOM06}
and \cite{Simeone-Somekh-Barness-Spagnolini-IT06}, respectively.
The basic framework employed in these works is the Wyner uplink
cellular model introduced in \cite{Wyner-94}. Following the linear
variant of this model, cells are arranged in a linear geometry and
only adjacent cells interfere with each other. Moreover,
inter-cell interference is described by a single parameter
$\alpha\in[0, 1]$, which defines the gain experienced by signals
travelling to interfered cells. Notwithstanding its simplicity,
this model captures the essential structure of a cellular system
and it provides insight into the system performance. The RTs added
to the basic Wyner model in
\cite{Simeone-Somekh-BarNess-Spagnolini-WCOM06}\cite{Simeone-Somekh-Barness-Spagnolini-IT06}
are assumed to operate in a half-duplex mode and to receive
signals from the MTs only (and not from adjacent RTs).

In this work we relax the latter restrictions by allowing
full-duplex operation at the RTs and considering the signal path
between adjacent RTs. Focusing on an intra-cell time-division
multiple-access (TDMA) operation and non-fading channels, we
assess the gain provided by the joint MCP approach over the
conventional single-cell processing (SCP) scheme by deriving the
per-cell sum-rate in the two scenarios. We finally remark that a
further contribution of this paper with respect to
\cite{Simeone-Somekh-BarNess-Spagnolini-WCOM06}\cite{Simeone-Somekh-Barness-Spagnolini-IT06}
is the extension to a relaying scenario of the analytical
framework introduced in \cite{Wyner-94}, whereby the signal
received by the BSs is interpreted as the outcome of a linear
time-invariant system.

\section{System Model}\begin{figure}[tb]
\begin{center}
\psfrag{Alpha\r}{\scriptsize$\alpha$}\psfrag{Beta\r}{\scriptsize$\beta$}
\psfrag{Gama\r}{\scriptsize$\gamma$}
\psfrag{Mu\r}{\scriptsize$\mu$}
\psfrag{Eta\r}{\scriptsize$\eta$}\psfrag{Xm\r}{\scriptsize$X_{m,n}$}
\psfrag{Xmm1\r}{\scriptsize$X_{m-1,n}$}\psfrag{Xmp1\r}{\scriptsize$X_{m+1,n}$}
\psfrag{Ym\r}{\scriptsize$Y_{m,n}$}\psfrag{Ymm1\r}{\scriptsize$Y_{m-1,n}$}\psfrag{Ymp1\r}{\scriptsize$Y_{m+1,n}$}
\psfrag{Rm\r}{\scriptsize$R_{m,n}$}\psfrag{Rmm1\r}{\scriptsize$R_{m-1,n}$}\psfrag{Rmp1\r}{\scriptsize$R_{m+1,n}$}
\psfrag{Wm\r}{\scriptsize$W_{m,n}$}\psfrag{Wmm1\r}{\scriptsize$W_{m-1,n}$}\psfrag{Wmp1\r}{\scriptsize$W_{m+1,n}$}
\psfrag{Zm\r}{\scriptsize$Z_{m,n}$}\psfrag{Zmm1\r}{\scriptsize$Z_{m-1,n}$}\psfrag{Zmp1\r}{\scriptsize$Z_{m+1,n}$}
\includegraphics[scale=0.5]{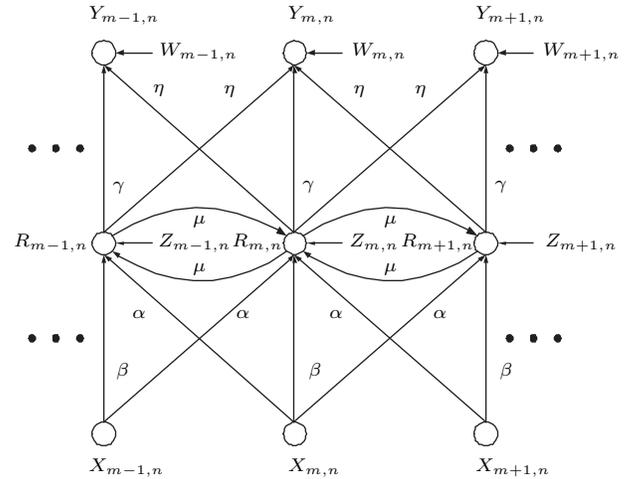}
\end{center}
\caption{Network model} \label{fig: Wyner relay network}
\end{figure}

We consider the uplink of a cellular system with a dedicated RT
for each transmitting MT. We focus on a scenario with no fading
and employ the framework of a linear cellular uplink channel
presented by Wyner \cite{Wyner-94}. RTs are added to the basic
Wyner model following the analysis in
\cite{Simeone-Somekh-BarNess-Spagnolini-WCOM06}\cite{Simeone-Somekh-Barness-Spagnolini-IT06}
(see Fig. \ref{fig: Wyner relay network} for a schematic diagram
of the setup). Throughout this paper we make the following
underlying assumptions:
\begin{itemize}
    \item The system includes infinitely many identical cells arranged on a
    line.
    \item A single MT is active in each cell at a given time (intra-cell TDMA protocol).
    \item A dedicated single RT is available in each cell to relay the signal from the MT.
    \item The signals from the MTs are received by the BSs via the relays (and not directly from the
    MTs).
    \item Each RT receives the signals of the MTs from its own cell and the two
    adjacent cells only.
    \item Each BS receives the signals of the RTs from its own cell and the two
    adjacent cells only.
    \item The channel power gain from the MT to its local RT,
    and its two adjacent RTs are denoted by $\beta^2$ and $\alpha^2$
    respectively.
    \item The channel power gain from the RT to its local BS,
    and its two adjacent BSs are denoted by $\eta^2$ and $\gamma^2$
    respectively.
    \item The channel power gain from the RT to its two adjacent RTs
    is $\mu^2$.
    \item The MTs use independent randomly generated complex Gaussian codebooks
    with zero mean and power $P$.
    \item The average transmit power of each RT is $Q$.
    \item The RTs are assumed to be oblivious and to use an AF relaying scheme.
    \item The RTs are assumed to be capable of receiving and
    transmitting simultaneously (i.e., we assume full-duplex operation, which amounts to assuming perfect echo-cancellation between transmit and receive paths).
    \item The RTs amplify and forward the received signal with a delay of
    $\lambda\ge 1$ symbols (an integer).
    \item The propagation delays between the different nodes of the
    system are negligible with respect to the symbol duration.
    \item No cooperation is assumed among MTs.
    \item No cooperation is assumed among RTs.
    \item All the attenuation parameters are known to the BSs.
\end{itemize}

The main differences between the current model and the model
presented in \cite{Simeone-Somekh-BarNess-Spagnolini-WCOM06}
\cite{Simeone-Somekh-Barness-Spagnolini-IT06}, are: (a)
full-duplex operation at the relays (which introduces the relaying
delay $\lambda$); (b) no direct connection between the MTs and the
BSs; and (c) the RTs receives also the signals of the two adjacent
MTs.

Accounting for the underlying assumptions listed above, a baseband
representation of the signal transmitted by the $m$'th RT for an
arbitrary time index $n$ is given by
\begin{multline}\label{eq: Relay received signal}
    R_{m,n}=g\left(\beta X_{m,n} + \alpha X_{m-1,n}+\alpha X_{m+1,n}+\right.\\
    \left.\mu R_{m-1,n-\lambda}+\mu R_{m+1,n-\lambda}+Z_{m,n}\right)\ ,
\end{multline}
where $Z$ represents the additive complex Gaussian noise process
$Z_{m,n}\sim\mathcal{CN}(0,\sigma^2_Z)$, which is assumed to be
independent and identically distributed (i.i.d.) with respect to
both the time and cell indices. The received signal at the $m$'th
BS antenna is given by
\begin{equation}\label{eq: BS received signal}
    Y_{m,n}=\gamma R_{m,n-\lambda} + \eta R_{m-1,n-\lambda}+\eta R_{m+1,n-\lambda}+W_{m,n}\ ,
\end{equation}
where $W$ represents the additive complex Gaussian noise process
$W_{m,n}\sim\mathcal{CN}(0,\sigma^2_W)$, which is assumed to be
i.i.d. with respect to both the time and cell indices and to be
statistically independent of $Z$. In addition, the RTs' gain $g$
is selected to satisfy the average power limitation
\begin{equation*}\label{eq: Relay power constraint}
    \sigma^2_r(g)\triangleq E\{\abs{R_{m,n}}^2\}\le Q\ .
\end{equation*}

\section{Sum-Rate Analysis}

In this section, we derive the per-cell sum-rate of the cellular
system at hand with MCP at the BSs and in the reference case with
SCP.

\subsection{Joint Multi-Cell Processing}
\begin{figure}[tb]
\begin{center}
\psfrag{H1\r}{$h_1$}\psfrag{H2\r}{$h_2$}\psfrag{Hr\r}{$h_r$}\psfrag{H3\r}{$h_3$}
\psfrag{X\r}{$X$} \psfrag{Y\r}{$Y$}
\psfrag{W\r}{$W$}\psfrag{Z\r}{$Z$} \psfrag{R\r}{$R$}
\includegraphics[scale=0.5]{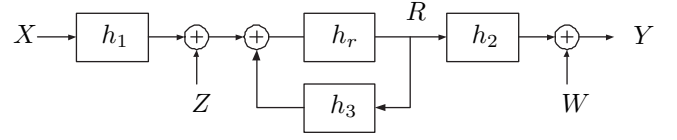}
\end{center}
\caption{Equivalent 2D LTI channel.} \label{fig: ISI channel}
\end{figure}

In this section we assume that the signals received at all BSs are
jointly decoded by an optimal central receiver. The receiver is
connected to the BSs via an ideal backbone and is assumed to be
aware of the Gaussian codebooks of all the MTs. It is noted that
using similar arguments as in \cite{Wyner-94}, it can be shown
that in this setup an intra-cell TDMA protocol is optimal.

Extending the one dimensional (1D) model introduced in
\cite{Wyner-94}, the linear equations \eqref{eq: Relay received
signal} and \eqref{eq: BS received signal} describing the network
of Fig. \ref{fig: Wyner relay network} can be interpreted as a two
dimensional (2D) linear time invariant (LTI) system. The block
diagram of the equivalent 2D LTI system is depicted in Fig.
\ref{fig: ISI channel} where the 2D filters read
\begin{equation}\label{eq: Filters "time" domain}
\begin{aligned}
    {h_1}_{m,n} &= \delta_n
    (\alpha\delta_{m-1}+\beta\delta_m+\alpha\delta_{m+1})\\
    {h_2}_{m,n} &=
    \delta_n(\eta\delta_{m-1}+\gamma\delta_m+\eta\delta_{m+1})\\
    {h_r}_{m,n} &= g\delta_{n-\lambda}\delta_m\\
    {h_3}_{m,n} &= \mu\delta_n(\delta_{m-1}+\delta_{m+1})\ ,
\end{aligned}
\end{equation}
with $\delta_n$ denoting the \emph{Kronecker} delta function. The
corresponding 2D \emph{Fourier} transforms of the signals in
\eqref{eq: Filters "time" domain} are given by
\begin{equation}\label{eq: Filters "frequency" domain}
\begin{aligned}
    H_1(\theta,\varphi) &=
    \beta+2\alpha \cos \theta\\
    H_2(\theta,\varphi) &=
    \gamma+2\eta\cos\theta\\
    H_r(\theta,\varphi) &= ge^{-j\lambda\varphi}\\
    H_3(\theta,\varphi) &= 2\mu\cos\theta\ .
\end{aligned}
\end{equation}
Since the noise processes $Z$ and $W$ are zero mean i.i.d. complex
Gaussian and statistically independent of each other and of the
input signal $X$, the output signal at the BSs can be expressed as
\begin{equation}\label{eq: ISI output signal}
    Y_{m,n}=S_{m,n}\ +\ N_{m,n}\ ,
\end{equation}
where $S_{m,n}$ and $N_{m,n}$ are zero mean wide sense stationary
(WSS) statistically independent processes representing the useful
part of the signal and the noise respectively. Now, using the 2D
extension of Szeg{\"o}'s theorem \cite{Wyner-94}, the achievable
rate in the channel \eqref{eq: ISI output signal} (without
spectral shaping), which is equal to the achievable per-cell
sum-rate of the network, is given for arbitrary $g$ by
\begin{equation}\label{eq: C AF implicit}
    R_{\mathrm{mcp}}=\frac{1}{(2\pi)^2}\int_0^{2\pi}\int_0^{2\pi}\log\left(1+\frac{\mathcal{S}_S(\theta,\varphi)}
    {\mathcal{S}_N(\theta,\varphi)}\right)d\varphi\ d\theta\ ,
\end{equation}
where $\mathcal{S}_S(\theta,\varphi)$ and
$\mathcal{S}_N(\theta,\varphi)$ are the 2D power spectral density
(PSD) functions of $S$ and $N$ respectively.

On examining Fig. \ref{fig: ISI channel}, we see that the PSD of
the useful signal is given by
\begin{equation}\label{eq: Spectrum S}
\mathcal{S}_S(\theta,\varphi)=P\abs{H_S(\theta,\varphi)}^2=P\abs{\frac{H_1
H_r H_2}{1-H_r H_3}}^2\ ,
\end{equation}
while the PSD of the noise is given by
\begin{equation}\label{eq: Spectrum N}
\mathcal{S}_N(\theta,\varphi)=\sigma_Z^2\abs{H_N(\theta,\varphi)}^2+\sigma_W^2=\sigma_Z^2\abs{\frac{H_r
H_2}{1-H_r H_3}}^2+\sigma_W^2\ ,
\end{equation}
where the transfer functions $H_1,\ H_2,\ H_r$, and $H_3$ are
defined in \eqref{eq: Filters "frequency" domain}.

\begin{prop}\label{prop: MCP AF sum-rate}
The per-cell sum-rate of MCP with AF relaying is given by
\begin{equation}\label{eq: MCP AF sum-rate}
R_{\mathrm{mcp}}=\frac{1}{2\pi}\int_0^{2\pi}\log\left(\frac{A+B+\sqrt{(A+B)^2-C^2}}{B+\sqrt{B^2-C^2}}\right)d\theta\
,
\end{equation}
where
\begin{equation*}\label{eq: A B C definitions}
\begin{aligned}
A&\triangleq P g^2(\beta+2\alpha\cos\theta)^2(\gamma+2\eta\cos\theta)^2\\
B&\triangleq\sigma_Z^2 g^2(\gamma+2\eta\cos\theta)^2+\sigma_W^2(1+4g^2\mu^2\cos^2\theta)\\
C&\triangleq 4\sigma_W^2 g\mu\cos\theta\ .
\end{aligned}
\end{equation*}
Furthermore, the optimal relay gain $g_\mathrm{o}$ is the unique
solution to the equation $\sigma^2_r(g)=Q$ where
\begin{equation}\label{eq: MCP relay power}
    \sigma^2_r(g)=\frac{(P\beta^2+\sigma^2_Z)g^2}{\sqrt{1-(2\mu
g)^4}}+\frac{4P\alpha^2 g^2}{\sqrt{1-(2\mu g)^2}+1-(2\mu g)^2}
\end{equation}
is the relay output power.
\end{prop}
\begin{proof}
See Appendix \ref{appx: MCP AF sum-rate}.
\end{proof}
It can be seen that the optimal gain is achieved when the relays
use their full power $Q$, and that
$g_\mathrm{o}\underset{Q\rightarrow\infty}{\longrightarrow}
1/(2\mu)$. Other observations are that the sum-rate
$R_{\mathrm{mcp}}$ is not interference limited and that it is
independent of the actual RT delay value $\lambda$. In the
following, we consider some relevant special cases.

\subsubsection{No adjacent RTs reception ($\mu=0$)}

This scenario refers to the case in which the RTs are employing
directional antennas pointed toward their local BSs (see also
discussion in \cite{Simeone-Somekh-BarNess-Spagnolini-WCOM06}
\cite{Simeone-Somekh-Barness-Spagnolini-IT06}). In this case, the
general expression \eqref{eq: MCP AF sum-rate} reduces to
\begin{multline}\label{eq: C AF explicit DA}
R_{\mathrm{mcp-da}}=\\
\frac{1}{2\pi}\int_0^{2\pi}\log \left(1+\frac{P
g^2(\beta+2\alpha\cos\theta)^2
(\gamma+2\eta\cos\theta)^2}{\sigma_Z^2g^2(\gamma+2\eta\cos\theta)^2+
\sigma_W^2}\right)d\theta\ .
\end{multline}
In addition, by setting $\mu=0$ in \eqref{eq: MCP relay power} we
obtain that
\begin{equation}\label{eq: Relay gain onstraint DA}
\begin{aligned}
    g^2_\mathrm{o}&= \frac{Q}{P(\beta^2+2\alpha^2)+\sigma^2_Z}\ .
\end{aligned}
\end{equation}

\subsubsection{Half-duplex operation}

In this case, the RTs are not capable of simultaneous
receive-transmit operation. Accordingly, the time is divided into
equal slots: during odd numbered slots the MTs are transmitting
with power $2P$ and the RTs only receive, while during even
numbered slots the MTs are silent and the RTs transmit. It is
easily verified that the per-cell sum-rate in this case is given
by multiplying \eqref{eq: C AF explicit DA} by $1/2$ while
replacing $P$ and $Q$ respectively with $2P$ and $2Q$, in both
\eqref{eq: C AF explicit DA} and \eqref{eq: Relay gain onstraint
DA}.

\subsection{Single Cell-Site Processing}

In this section we consider a conventional SCP scheme in which no
cooperation between cells is allowed. According to this scheme,
each cell-site receiver is aware of the codebooks of its own users
only, and it treats all other cell-site signals as interference.
Notice that since the RTs are oblivious, their AF operation is not
influenced by the fact that the BSs are not cooperating. In
addition, since the input signals and noise statistics remain the
same, expression \eqref{eq: MCP relay power} is also valid for the
current setup.

The output signal can be expressed as
\begin{equation*}\label{eq: SCP output signal}
    Y_{m,n}={S_U}_{m,n} +{S_I}_{m,n}+N_{m,n}\ ,
\end{equation*}
where the useful part of the output signal $S_U$ is defined as
\begin{equation*}\label{eq: SCP usful signal def}
{S_U}_{m,n} = \sum_{l=-\infty}^{\infty}{h_S}_{0,n-l}X_{m,l}\ ,
\end{equation*}
and $h_S$ and $h_N$ are the signal and noise space-time impulse
response functions whose Fourier transforms are given in
\eqref{eq: Spectrum S} and \eqref{eq: Spectrum N} respectively.
The interference part of the output signal $S_I$ is defined as
\begin{equation*}\label{eq: SCP intereference signal def}
{S_I}_{m,n} = \sum_{ \underset{l_1\neq
m}{l_1=-\infty}}^{\infty}\sum_{l_2=-\infty}^{\infty}{h_S}_{m-l_1,n-l_2}X_{l_1,l_2}\
,
\end{equation*}
and the noise part of the signal is defined as
\begin{equation*}\label{eq: SCP noise def}
    N_{m,n}= \sum_{l_1=-\infty}^{\infty}\sum_{l_2=-\infty}^{\infty}{h_N}_{m-l_1,n-l_2}Z_{l_1,l_2}+
    W_{m,n}\ .
\end{equation*}
Since $X$, $Z$, and $W$ are independent of each other, zero-mean
complex Gaussian and i.i.d. in space and time, it is easily
verified that $S_U$, $S_I$, and $N$ are independent and zero-mean
complex Gaussian as well. It is also evident that for each $m$ the
processes are WSS along the time axis $n$. Accordingly, the output
process at the $m$'th cell can be seen as a Gaussian inter-symbol
interference (ISI) channel with additive colored independent
interference and noise.
\begin{prop}\label{prop: SCP sum-rate}
The per-cell sum-rate of SCP with AF relaying is given for an
arbitrary relay gain $0<g<g_\mathrm{o}$, by
\begin{equation*}\label{eq: SCP sum-rate}
R_{\mathrm{scp}}=\frac{1}{2\pi}\int_0^{2\pi}\log\left(1+
\frac{\mathcal{S}_U(\varphi)}{\mathcal{S}_I(\varphi)+\mathcal{S}_N(\varphi)}\right)d\varphi\
,
\end{equation*}
where $\mathcal{S}_U(\varphi)$, $\mathcal{S}_I(\varphi)$, and
$\mathcal{S}_N(\varphi)$ are the 1D PSDs of the useful signal,
interference, and noise respectively:
\begin{equation*}\label{eq: SCP PSD explicit}
\begin{aligned}
\mathcal{S}_U(\varphi)
&=\frac{P}{(2\pi)^2}\abs{\int_0^{2\pi}H_S(\theta,\varphi)d\theta}^2\\
\mathcal{S}_I(\varphi)
&=\frac{P}{2\pi}\int_0^{2\pi}\abs{H_S(\theta,\varphi)}^2 d\theta-
\frac{P}{(2\pi)^2}\abs{\int_0^{2\pi}H_S(\theta,\varphi)d\theta}^2\\
\mathcal{S}_N(\varphi) &= \frac{\sigma^2_Z}{2\pi}\int_0^{2\pi}
\abs{H_N(\theta,\varphi)}^2 d\theta + \sigma^2_W\ .
\end{aligned}
\end{equation*}
\end{prop}
\begin{proof}
See Appendix \ref{appx: SCP sum-rate}.
\end{proof}
It is noted that in contrast to the MCP scheme, $R_{\mathrm{scp}}$
is interference limited. It is also easy to verify that
$R_{\mathrm{scp}}$ is independent of the actual RT delay value
$\lambda$.

\section{Numerical Results}

\begin{figure}[tb]
\begin{center}
\includegraphics[scale=0.5]{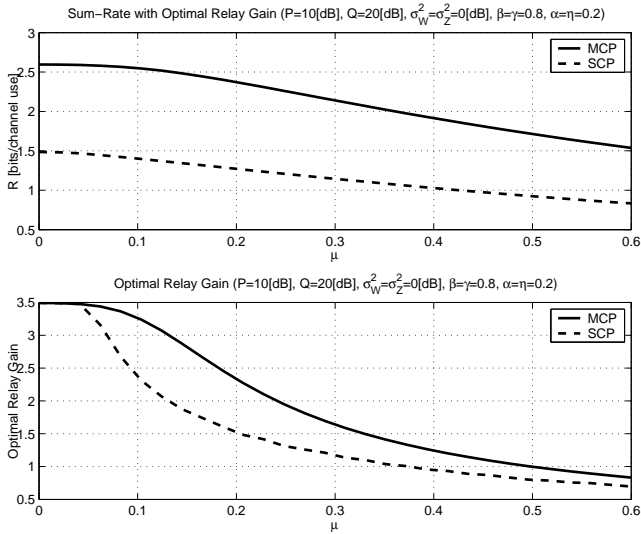}
\end{center}
\caption{Sum-rates per-cell} \label{fig: Sum-rates}
\end{figure}

In Fig. \ref{fig: Sum-rates}-a the sum-rates per-cell of the MCP
and the SCP schemes are plotted as functions of the inter-relay
interference factor $\mu$ for $P/\sigma^2=10$ [dB], $Q/\sigma^2\le
20$ [dB], $\sigma^2_Z=\sigma^2_W=\sigma^2=1$, $\alpha=\eta=0.2$,
and $\beta=\gamma=0.8$. The curves are plotted for an optimal
selection of the relay gain $g$, which is shown for both schemes
in Fig. \ref{fig: Sum-rates}-b. Examining the figures, it is
observed that for this setting the MCP scheme demonstrates a
meaningful improvement on performance over the SCP scheme. The
deleterious effect of increasing inter-relay interference $\mu$ is
also demonstrated for both schemes. Moreover, the optimal relay
gain for both schemes also decreases with $\mu$. Another
observation is that the optimal gain of the SCP scheme is lower
than that of the of the MCP scheme for $\mu$ larger than some
threshold. Hence, using the full power of the RTs is sub-optimal
for the SCP scheme under certain conditions.

\section{Concluding Remarks}

In this paper, joint MCP of MTs that are received only via
dedicated RTs applying full-duplex AF relaying, has been
considered. The received signal at the BSs can be seen as the
output of a 2D LTI channel. Using the 2D version of Szeg{\"o}'s
Theorem, a closed form expression for the achievable per-cell
sum-rate of intra-cell TDMA protocol has been derived. As a
reference the rate of a conventional SCP scheme, which treats
other cell MTs' signals as interference, has also been derived.
Comparing the rates of the two schemes, the benefits of the MCP
scheme has been demonstrated. Moreover, we have observed that the
rates of both schemes are decreasing with the intra-relay
interference factor, $\mu$. The latter can be explained for the
MCP scheme, by the fact that the equivalent 2D LTI channel becomes
more distorted with increasing $\mu$. Since no MTs cooperation is
allowed and no rate splitting is used, this distortion can not be
mitigated by power allocation over time or space, and the
resulting rate decreases with $\mu$. We also have shown that using
the full power of the RTs is unconditionally optimal only for the
MCP scheme. Numerical results have revealed that under certain
conditions, the SCP setting produces an equivalent noisy ISI
channel, the rate of which is not necessarily maximized by using
the full RTs power. Other more sophisticated relaying schemes, are
currently under further investigation.

\appendix

\subsection{Proof of Proposition \ref{prop: MCP AF sum-rate}}\label{appx: MCP AF sum-rate}
It is easily verified that the RT output signal $R_{m,n}$
\eqref{eq: Relay received signal} is a WSS complex Gaussian 2D
process with zero mean. Hence, its power can be expressed by
\begin{equation}\label{eq: Relay gain onstraint}
\begin{aligned}
    &\sigma^2_r(g)=E\{\abs{R_{m,n}}^2\}\\
    &=\frac{1}{(2\pi)^2}\int_0^{2\pi}\int_0^{2\pi}\frac{(P\abs{H_1}^2+\sigma^2_Z)\abs{H_r}^2}{\abs{1-H_r
    H_3}^2}\ d\varphi\ d\theta\\
    &=\int_0^{2\pi}\int_0^{2\pi}\frac{(2\pi)^{-2}(P(\beta+2\alpha\cos\theta)^2
    +\sigma^2_Z)g^2}{1-4g\mu\cos\theta\cos(\lambda\varphi)+4g^2\mu^2\cos^2\theta}\ d\varphi\ d\theta\ ,
\end{aligned}
\end{equation}
where the third equality is achieved by substituting \eqref{eq:
Filters "frequency" domain}. Examining \eqref{eq: Relay gain
onstraint}, it is clear that in order for the relay to transmit
finite power (or for the whole system to be stable) the poles of
the integrand must lie inside the unit circle. Assuming that $g$
is real this condition implies that
\begin{equation*}\label{eq: Condition}
    g\ <\ \frac{1}{2\mu}\ .
\end{equation*}
It is also verified by differentiating the integrand of \eqref{eq:
Relay gain onstraint} with respect to $g$ that $\sigma_r^2(g)$ is
an increasing function of $g$ with $\sigma_r^2(0)=0$. By making a
change of variable $\varphi'=\lambda\varphi$, and integrating
\eqref{eq: Relay gain onstraint} over $\varphi'$ we get
\begin{equation}\label{eq: Relay gain onstraint single integration}
\begin{aligned}
\sigma^2_r(g)&=\frac{1}{2\pi}\int_0^{2\pi}\frac{(P(\beta+2\alpha\cos\theta)^2
    +\sigma^2_Z)g^2}{1-4g^2\mu^2\cos^2\theta}\ d\theta\ ,
\end{aligned}
\end{equation}
where the last equality is achieved by using formula 3.616.2 of
\cite{Gradshteyn-Ryzhik-6th} and some algebra. It is noted that
\eqref{eq: Relay gain onstraint single integration} implies that
the power of the relay signal is \emph{independent} of the actual
relay delay duration. Expression \eqref{eq: Relay gain onstraint
single integration} can be further simplified into its final
closed form of \eqref{eq: MCP relay power}, by applying formulas
3.653.2 and 3.682.2 of \cite{Gradshteyn-Ryzhik-6th} and some
additional algebra.

To derive the per-cell sum-rate expression for an arbitrary RT
gain $g$, we substitute \eqref{eq: Spectrum S} and \eqref{eq:
Spectrum N} into \eqref{eq: C AF implicit} to obtain
\begin{multline}\label{eq: C AF explicit}
R_{\mathrm{mcp}}=\frac{1}{(2\pi)^2}\int_0^{2\pi}\int_0^{2\pi}\\
\log\left(1+\frac{P\abs{H_1 H_r H_2}^2}{\sigma_Z^2\abs{H_r
H_2}^2+\sigma_W^2\abs{1-H_r H_3}^2}\right)d\varphi\ d\theta\ .
\end{multline}
It is easily verified by differentiating the integrand of
\eqref{eq: C AF explicit} with respect to $g$, that the rate is an
increasing function of the RT gain $g$ for $0\le g <1/(2\mu)$. We
can conclude that, since $\sigma^2_r(g)$ is also an increasing
function of $g$, the rate is maximized when the RTs use their full
power by setting their gain to $g_o$ which is the unique solution
to $\sigma^2_r(g)=Q$. Finally, by substituting \eqref{eq: Filters
"frequency" domain}, applying formula 4.224.9 of
\cite{Gradshteyn-Ryzhik-6th} twice to \eqref{eq: C AF explicit},
and using some algebra we obtain \eqref{eq: MCP AF sum-rate}.

\subsection{Proof of Proposition \ref{prop: SCP sum-rate}}\label{appx: SCP sum-rate}
First, we express the three PSDs of interest in terms of the
system signal and noise 2D transfer functions
$H_S(\theta,\varphi)$ and $H_N(\theta,\varphi)$. Starting with the
noise component, it is easily verified that its PSD is given by
\begin{equation*}\label{eq: SCP noise PSD}
\begin{aligned}
\mathcal{S}_N(\varphi) &= \frac{1}{2\pi}\int_0^{2\pi}
\mathcal{S}_N(\theta,\varphi) d\theta\\
&= \sigma^2_Z\frac{1}{2\pi}\int_0^{2\pi}
\abs{H_N(\theta,\varphi)}^2 d\theta + \sigma^2_W\ ,
\end{aligned}
\end{equation*}
where the 2D filter $H_N(\theta,\varphi)$ is defined in \eqref{eq:
Spectrum N}.

To calculate the useful signal PSD, let us define the following 2D
filter ${\hat{h}_U}{}_{m,n}\triangleq \delta_m {h_S}_{m,n}$\ .
It is easily verified that
\begin{equation*}\label{eq: SCP usful signal with aux filter}
{S_U}_{m,n} =
\sum_{l_1=-\infty}^{\infty}\sum_{l_2=-\infty}^{\infty}{\hat{h}_U}{}_{l_1-m,l_2-n}X_{l_1,l_2}\
,
\end{equation*}
and that the 2D Fourier transform of ${\hat{h}_U}{}_{m,n}$ is
given by
\begin{equation*}\label{eq: SCP usful aux filter transform}
\begin{aligned}
\hat{H}_U(\theta,\varphi)&=\mathcal{F}\{{h_S}_{m,n}\}\ast\ast\mathcal{F}\{\delta_m\}
=H_S(\theta,\varphi)\ast\ast 2\pi\delta(\varphi)\\
&=\frac{1}{2\pi}\int_0^{2\pi}H_S(\theta,\varphi)d\theta\ ,
\end{aligned}
\end{equation*}
where $\ast\ast$ denotes a 2D cyclic convolution operation, and
$\delta(\varphi)$ denotes the \emph{Dirac} delta function. Hence,
the useful signal PSD becomes
\begin{equation*}\label{eq: SCP useful PSD explicit}
\begin{aligned}
\mathcal{S}_U(\varphi) &= P
\frac{1}{2\pi}\int_0^{2\pi}\abs{\hat{H}_U(\theta,\varphi)}^2
d\theta\\
&=P\frac{1}{2\pi}\int_0^{2\pi}\abs{\frac{1}{2\pi}\int_0^{2\pi}H_S(\theta',\varphi)d\theta'}^2
d\theta\\
&=P\frac{1}{(2\pi)^2}\abs{\int_0^{2\pi}H_S(\theta,\varphi)d\theta}^2\
.
\end{aligned}
\end{equation*}

To calculate the interference PSD, let us define the following 2D
filter ${\hat{h}_I}{}_{m,n}\triangleq (1-\delta_m) {h_S}_{m,n}$\ .
Then we have that
\begin{equation*}\label{eq: SCP interference with aux filter}
{S_I}_{m,n} =
\sum_{l_1=-\infty}^{\infty}\sum_{l_2=-\infty}^{\infty}{\hat{h}_I}{}_{l_1-m,l_2-n}X_{l_1,l_2}\
,
\end{equation*}
and that the 2D Fourier transform of ${\hat{h}_S}{}_{m,n}$ is
given by
\begin{equation*}\label{eq: SCP interference aux filter transform}
\begin{aligned}
\hat{H}_I(\theta,\varphi)&=\mathcal{F}\{{h_S}_{m,n}\}\ast\ast\mathcal{F}\{1-\delta_m\}\\
&=H_S(\theta,\varphi)\ast\ast ((2\pi)^2\delta(\theta)\delta(\varphi)-2\pi\delta(\varphi))\\
&=H_S(\theta,\varphi)-\frac{1}{2\pi}\int_0^{2\pi}H_S(\theta,\varphi)d\theta\
.
\end{aligned}
\end{equation*}
Hence, the interference PSD is given by
\begin{equation*}\label{eq: SCP interference PSD explicit}
\begin{aligned}
&\mathcal{S}_I(\varphi) = P
\frac{1}{2\pi}\int_0^{2\pi}\abs{\hat{H}_I(\theta,\varphi)}^2
d\theta\\
&=P\frac{1}{2\pi}\int_0^{2\pi}\abs{H_S(\theta,\varphi)-\frac{1}{2\pi}\int_0^{2\pi}H_S(\theta',\varphi)d\theta'}^2
d\theta\\
&=P\frac{1}{2\pi}\int_0^{2\pi}\abs{H_S(\theta,\varphi)}^2 d\theta-
P\frac{1}{(2\pi)^2}\abs{\int_0^{2\pi}H_S(\theta,\varphi)d\theta}^2\
.
\end{aligned}
\end{equation*}

\section*{Acknowledgment}
The research was supported by a Marie Curie Outgoing International
Fellowship and the NEWCOM network of excellence both within the
6th European Community Framework Programme, and by the U.S.
National Science Foundation under Grants ANI-03-38807 and
CNS-06-25637.



\begin{thebibliography}{10}

\bibitem{Simeone-Somekh-BarNess-Spagnolini-WCOM06}
O.~Simeone, O.~Somekh, Y.~Bar-Ness, and U.~Spagnolini, ``Uplink
throughput of {TDMA} cellular systems with multicell processing
and amplify-and-forward cooperation between mobiles,'' {\em IEEE
Trans. Wireless Commun.}, to appear.

\bibitem{Simeone-Somekh-Barness-Spagnolini-IT06}
O.~Simeone, O.~Somekh, Y.~Bar-Ness, and U.~Spagnolini,
``Throughput of low-power cellular systems with collaboration at
base stations and mobile terminals.'' Submitted to \emph{IEEE
Trans. Inform. Theory}, 2006.

\bibitem{Lin-Hsu-Infocom00}
Y.-D.~J. Lin and Y.-C. Hsu, ``Multihop cellular: A new
architecture for wireless communications,'' in {\em Proc. IEEE
{INFOCOM} (3)}, (Tel-Aviv, Israel), pp.~1273--1282, Mar. 26--30,
2000.

\bibitem{Pabst-Walke-Schultz-Herhold-Yanikomeroglu-Mukherjee-Viswanathan-Lott-%
Sirwas-Falconer-Fettweis-COMMAG04} R.~Pabst \emph{et-al.},
``Relay-based deployment concepts for wireless and mobile
broadband radio,'' {\em IEEE Commun. Mag.}, pp.~80--89, Sep. 2004.

\bibitem{Zemlianov-Veciana-JSAC05}
A.~Zemlianov and G.~{de Veciana}, ``Capacity of ad hoc wireless
network with infrastructure support,'' {\em IEEE Journal on
Selected Areas in Commun.}, vol.~23, pp.~657--667, Mar. 2005.

\bibitem{Liu-Liu-Towsely-INFOCOM03}
B.~Liu, Z.~Liu, and D.~Towsely, ``On the capacity of hybrid
wireless networks,'' in {\em Proc. IEEE INFOCOM}, (San-Francisco
CA, USA), pp.~1543--1552, Mar. 30--Apr. 3, 2003.

\bibitem{Kramer-Gastpar-Gupta-IT05}
G.~Kramer, M.~Gastpar, and P.~Gupta, ``Cooperative strategies and
capacity theorems for relay networks,'' {\em IEEE Trans. Inform.
Theory}, vol.~51, pp.~3037--3063, Sep. 2005.

\bibitem{Zhou-Zhao-Xu-Yao-COMMAG03}
S.~Zhou, M.~Zhao, X.~Xu, and Y.~Yao, ``Distributed wireless
communication system: A new architecture for public wireless
access,'' {\em IEEE Commun. Mag.}, pp.~108--113, Mar. 2003.

\bibitem{Wyner-94}
A.~D. Wyner, ``Shannon-theoretic approach to a {G}aussian cellular
  multiple-access channel,'' {\em IEEE Trans. Inform.
Theory}, vol.~40, pp.~1713--1727, Nov. 1994.

\bibitem{Hanly-Whiting-Telc-1993}
S.~V. Hanly and P.~A. Whiting, ``Information-theoretic capacity of
  multi-receiver networks,'' {\em Telecommun. Syst.}, vol.~1, pp.~1--42, 1993.

\bibitem{Somekh-Shamai-2000}
O.~Somekh and S.~{Shamai (Shitz)}, ``Shannon-theoretic approach to
a {G}aussian cellular multi-access channel with fading,'' {\em
IEEE Trans. Inform. Theory}, vol.~46, pp.~1401--1425, Jul. 2000.

\bibitem{Shamai-Somekh-Zaidel-JWCC-2004}
S.~{Shamai (Shitz)}, O.~Somekh, and B.~M. Zaidel, ``Multi-cell
communications: An information theoretic perspective,'' in {\em
Proc. of the Joint Workshop on Commun. and Coding (JWCC'04)},
(Donnini, Florence, Italy), Oct.14--17, 2004.

\bibitem{Somekh-Simeone-Barness-Haimovich-Shamai-BookChapt-07}
O.~Somekh, O.~Simeone, Y.~{Bar-Ness}, A.~M. Haimovich,
U.~Spagnolini, and S.~{Shamai (Shitz)}, {\em Distributed Antenna
Systems: Open Architecture for Future Wireless Communications},
ch.~An Information Theoretic View of Distributed Antenna
Processing in Cellular Systems. \newblock Auerbach Publications,
CRC Press, May 2007.

\bibitem{Gradshteyn-Ryzhik-6th}
I.~S. Gradshteyn and I.~M. Ryzhik, {\em Table of Integrals,
Series, and Products}. \newblock Academic Press, 6~ed., 2000.

\end{thebibliography}

\end{document}